\documentclass{ws-rv9x6}
\usepackage{subfigure}   
\usepackage{ws-rv-van}   % numbered citation & references (default)
\makeindex
\begin{document}

\chapter[Fluctuations and correlations in reaction kinetics and population
            dynamics]{Fluctuations and Correlations in Chemical Reaction \\
            Kinetics and Population Dynamics} 
\label{uct}

\author[U.~C. T\"auber]{Uwe C. T\"auber} 
%\footnote{Author footnote.}} %\index[aindx]{T\"auber, U.C.} 
\address{Department of Physics (MC 0435) and \\
             Center for Soft Matter and Biological Physics, Virginia Tech, \\
		850 West Campus Drive, Blacksburg, VA 24061, USA \\ 
             tauber@vt.edu} %\footnote{Virginia Tech.}}

\begin{abstract}
This chapter provides a pedagogical introduction and overview of spatial and
temporal correlation and fluctuation effects resulting from the fundamentally
stochastic kinetics underlying chemical reactions and the dynamics of
populations or epidemics. 
After reviewing the assumptions and mean-field type approximations involved
in the construction of chemical rate equations for uniform reactant densities,
we first discuss spatial clustering in birth-death systems, where non-linearities
are introduced through either density-limiting pair reactions, or equivalently
via local imposition of finite carrying capacities.
The competition of offspring production, death, and non-linear inhibition
induces a population extinction threshold, which represents a non-equilibrium
phase transition that separates active from absorbing states.
This continuous transition is characterized by the universal scaling exponents
of critical directed percolation clusters.
Next we focus on the emergence of depletion zones in single-species
annihilation processes and spatial population segregation with the associated
reaction fronts in two-species pair annihilation. 
These strong (anti-)correlation effects are dynamically generated by the
underlying stochastic kinetics. 
Finally, we address noise-induced and fluctuation-stabilized spatio-temporal
patterns in basic predator-prey systems, exemplified by spreading activity
fronts in the two-species Lotka--Volterra model as well as spiral structures in
the May--Leonard variant of cyclically competing three-species systems akin to
rock-paper-scissors games.
\end{abstract}

%\markright{Customized Running Head for Odd Page} 
\body

% \newpage \tableofcontents

\section{Introduction}

The kinetics of chemical reactions, wherein the identity or number of reactant
particles changes either spontaneously or upon encounter, constitutes a 
highly active research field in non-equilibrium statistical physics of 
stochastically interacting particle systems, owing both to the fundamental 
questions it addresses as well as its broad range of applications.
Of specific interest are reaction-diffusion models that for example capture 
chemical reactions on catalytic solid surfaces or in gels where convective 
transport is inhibited.
This scenario of course naturally pertains to genuine reactions in chemistry or biochemistry, and in nuclear, astro-, and particle physics.
Yet reaction-diffusion models are in addition widely utilized for the
quantitative description of a rich variety of phenomena in quite distinct
disciplines that range from population dynamics in ecology, growth and 
competition of bacterial colonies in microbiology, the dynamics of topological 
defects in the early universe in cosmology, equity and financial markets in 
economics, opinion exchange and the formation of segregated society 
factions in sociology, and many more.
Reactive `particles' also emerge as relevant effective degrees of freedom in
other physical applications; prominent examples include excitons kinetics in 
organic semiconductors, domain wall interactions in magnets, and interface 
dynamics in stochastic growth models. 

The traditional textbook literature, e.g., in physical chemistry and 
mathematical biology, almost exclusively focuses on a description of reacting 
particle systems in terms of coupled deterministic non-linear rate equations.
While these certainly represent an indispensable tool to characterize such
complex dynamical systems, they are ultimately based on certain mean-field
approximations, usually involving the factorization of higher moments of
stochastic variables in terms of products of their means, as in the classical
Guldberg--Waage law of mass action.
Consequently, both temporal and spatial correlations are neglected in such
treatments, and are in fact also only rudimentarily captured in standard
spatial extensions of the mean-field rate equations to reaction-diffusion 
models.
Under non-equilibrium conditions, however, spatio-temporal fluctuations
often play a quite significant role and may even qualitatively modify the
dynamics, as has been clearly established over the past decades through
extensive Monte Carlo computer simulations, a remarkable series of exact
mathematical treatments (albeit mostly in one dimension), and a variety of
insightful approximative analytical schemes that extend beyond mean-field
theory.
These include mappings of the stochastic dynamics to effective field theories 
and subsequent analysis by means of renormalization group methods.

In the following, we shall discuss the eminent influence of spatio-temporal
fluctuations and self-generated correlations in five rather simple particle 
reaction and, in the same general language, population ecology models
that however display intriguing non-trivial dynamical features:

(1) As in thermal equilibrium, strong fluctuations and long-range correlations 
emerge in the vicinity of continuous phase transitions between distinct
non-equilibrium steady states, as we shall exemplify for the extinction
threshold that separates and active from an absorbing state.
In this paradigmatic situation, the dynamical critical properties are described
by the scaling exponents of critical directed percolation clusters which assume
non-mean-field values below the critical dimension $d_c = 4$.

(2) Crucial spatio-temporal correlations may also be generated by the
chemical kinetics itself; this is indeed the case for simple single-species pair 
annihilation reactions that produce long-lived depletion zones in dimensions 
$d \leq 2$ which in turn slow down the resulting algebraic density decay.

(3) For two-species binary annihilation, the physics becomes even richer: for
$d \leq 4$, particle anti-correlations induce species segregation into
chemically inert, growing domains, with the reactions confined to their
interfaces.

(4) Spatially extended stochastic variants of the classical Lotka--Volterra 
model for predator-prey competition and coexistence display remarkably rich 
noise-generated and -stabilized dynamical structures, namely spreading
activity fronts that lead to erratic but persistent population oscillations.

(5) Cyclic competition models akin to the rock-paper-scissors game too 
produce intriguing spatio-temporal structures, whose shape is determined
in a subtle manner by the presence or absence of conservation laws in the
stochastic dynamics: 
Three species subject to cyclic Lotka--Volterra competition with conserved
total particle number organize into fluctuating clusters, whereas characteristic
spiral patterns form in the May--Leonard model with distinct predation and 
birth reactions.

\section{Chemical Master and Rate Equations}

\subsection{Stochastic reaction processes}

To begin, we consider simple \emph{death-birth} reactions \cite{hh83, hs98,
jdm02, krb10, uct14}, with reactants of a single species $A$ either
spontaneously decaying away or irreversibly reaching a chemically inert state
$\emptyset$: $A \to \emptyset$ with rate $\mu$; or producing identical
offspring particles, e.g.: $A \to A + A$ with rate $\sigma$.
Note that we may also view species $A$ as indicating individuals afflicted with
a contagious disease from which they may recover with rate $\mu$ or that 
they can spread among others with rate $\sigma$.
We shall consider these reactions as continuous-time Markovian stochastic
processes that are fully determined by prescribing the transition rates from
any given system configuration at instant $t$ to an infinitesimally later time 
$t + dt$.
Assuming mere local processes, i.e., for now ignoring any spatial degrees of
freedom, our reaction model is fully characterized by specifying the number
$n$ of particles or individuals of species $A$ at time $t$. 
The death-birth reactions are then encoded in the transition rates
$w(n \to n - 1) = \mu \, n$ and $w(n \to n+1) = \sigma \, n$ that linearly
depend on the instantaneous particle number $n$.
Accounting for both gain and loss terms for the configurational probability
$P(n,t)$ then immediately yields the chemical \emph{master equation}
\cite{vk81, hh83, krb10, vv10, uct14}
\begin{equation}
  \frac{\partial P(n,t)}{\partial t} \bigg\vert_\mathrm{db} = 
  \mu \, (n+1) \, P(n+1,t) - \left( \mu + \sigma \right) n \, P(n,t) 
  + \sigma \, (n-1) \, P(n-1,t) \ .
\label{masdb}
\end{equation}

This temporal evolution of the probability $P(n,t)$ directly transfers to its
moments
$\left\langle n(t)^k \right\rangle = \sum_{n=0}^\infty n^k \, P(n,t)$.
For example, for the \emph{mean particle number} 
$a(t) = \langle n(t) \rangle$, a straightforward summation index shift results
in the exact linear differential equation
\begin{eqnarray}
  &&\frac{\partial a(t)}{\partial t} \bigg\vert_\mathrm{db} =
  \sum_{n=1}^\infty n \, \frac{\partial P(n,t)}{\partial t}
  \bigg\vert_\mathrm{db} 
\label{reqdb} \\ 
  &&\qquad = \sum_{n=1}^\infty \left[ \mu \, n (n-1) 
  - \left( \mu + \sigma \right) n^2 + \sigma \, n (n+1) \right] P(n,t) = 
  \left( \sigma - \mu \right) a(t) \ , \nonumber
\end{eqnarray}
as the terms $\sim n^2$ in the bracket all cancel. 
Its solution $a(t) = a(0) \, e^{\left( \sigma - \mu \right) t}$ naturally 
indicates that if the particle decay rate is faster than the birth rate,
$\mu > \sigma$,  the population will go extinct and reach the inactive,
\emph{absorbing} empty state $a = 0$, whereupon all reactions irretrievably
cease.
In stark contrast, Malthusian exponential population explosion ensues for 
$\sigma > \mu$.

In order to prevent an unrealistic population divergence, one may impose a
non-linear process that effectively limits the reactant number; for example,
we could add coagulation $A + A \to A$ with reaction rate $\lambda$ that
can be viewed as mimicking constraints imposed by locally restricted
resources.
In the context of disease spreading, this scenario is often referred to as
\emph{simple epidemic process}.
The frequency of such binary processes annihilating one of the reactants is 
proportional to the number of particle pairs in the system, whence the
associated transition rate becomes $w(n \to n-1) = \lambda \, n (n-1)$, and
the master equation reads \cite{kk88, otb89, krb10, uct14}
\begin{equation}
  \frac{\partial P(n,t)}{\partial t} \bigg\vert_\mathrm{an} = 
  \lambda \left[ n (n+1) \, P(n+1,t) - n (n-1) \, P(n,t) \right] \, .
\label{masan}
\end{equation}
Proceeding as before, one now finds for the mean particle number decay
\begin{equation}
  \frac{\partial a(t)}{\partial t} \bigg\vert_\mathrm{an} = \lambda
  \sum_{n=1}^\infty \left[ n (n-1)^2 - n^2 (n-1) \right] P(n,t) = 
  - \lambda \left\langle [n (n-1)](t) \right\rangle \, . 
\label{exqpa}
\end{equation}
As to be expected, it is governed by the instantaneous number of particle
pairs; that quantity involves the second moment $\langle n(t)^2 \rangle$, 
whose time evolution in turn is determined by the third moment, etc.
Consequently, one faces an infinite hierarchy of moment differential equations
that is much more difficult to analyze than the simple closed \eref{reqdb}.

\subsection{Mean-field rate equation approximation}

A commonly applied scheme to close the moment hierarchy for non-linear
stochastic processes is to impose a mean-field type factorization for higher
moments.
The simplest such approximation entails neglecting any fluctuations, setting
the connected two-point correlation function to zero,
$C(t,t') = \langle n(t) \, n(t') \rangle - \langle n(t) \rangle \langle n(t') \rangle
\approx 0$.
This assumption should hold best for large populations $n \gg 1$, when 
relative mean-square fluctuations $(\Delta n)^2 / \langle n \rangle^2 =
C(t,t) / a(t)^2$ should be small; \eref{exqpa} then simplifies to the kinetic
\emph{rate equation}
\begin{equation}
  \frac{\partial a(t)}{\partial t} \bigg\vert_\mathrm{an} \approx 
  - \lambda \, a(t)^2 \ , 
\label{reqpa}
\end{equation}
or $\partial a(t)^{-1} / \partial t \approx \lambda$.
It is readily integrated to $a(t)^{-1} = a(0)^{-1} + \lambda \, t$, i.e.,
\begin{equation}
  a(t)  = \frac{a(0)}{1 + \lambda \, a(0) \, t} \ ,
\label{mfspa}
\end{equation}
which becomes independent of the initial particle number $a(0)$ and decays 
to zero algebraically $\sim 1 / \lambda \, t$ for large times 
$t \gg 1 / \lambda \, a(0)$.
Note that the right-hand side of the rate equations resulting from mean-field
factorizations of non-linear reaction terms encode the corresponding
stochiometric numbers as powers of the reactant numbers, precisely as in the
Guldberg--Waage law of mass action describing reaction concentration
products in chemical equilibrium \cite{vv10, uct14}.
Already in non-spatial systems, these factorizations disregard any temporal
fluctuations; in spatially extended systems, they moreover assume well-mixed
and thus homogeneously distributed reactants.

\section{Population Dynamics with Finite Carrying Capacity}

\subsection{Mean-field rate equation analysis}

Next we combine the three reaction processes of the preceding section to
arrive at the simplest possible population dynamics model for a single species
that incorporates death with rate $\mu$, (asexual) reproduction with rate
$\sigma$, and (non-linear) competition with rate $\lambda$ to constrain the
active-state particle number \cite{hs98, jdm02}.
Adding the right-hand sides of Eqs.~(\ref{reqdb}) and (\ref{reqpa}) yields
the associated mean-field rate equation for the particle number or population:
\begin{equation}
  \frac{\partial a(t)}{\partial t}  \approx \left( \sigma - \mu \right) a(t) 
  - \lambda \, a(t)^2 = \lambda \, a(t) \left[ r - a(t) \right] \, .
\label{reqdp}
\end{equation}
In the last step we have cast the rate equation in the form of a \emph{logistic
model} with \emph{carrying capacity} 
$r = \left( \sigma - \mu \right) / \lambda$.
Its stationary solutions are the absorbing state $a = 0$ and a population
number equal to the carrying capacity $a = r$.
Straightforward linear stability analysis of \eref{reqdp} establishes that the 
latter stationary state is approached for $\sigma > \mu$ or $r > 0$, whereas
of course $a(t) \to 0$ for $\sigma < \mu$ ($r < 0$). 

For $\sigma \not= \mu$, the ordinary first-order differential equation 
(\ref{reqdp}) is readily integrated after variable separation:
\begin{equation*}
  \lambda \, t = \frac{1}{r} \int_{a(0)}^{a(t)} 
  \left( \frac{1}{a} +\frac{1}{r - a} \right) da = \frac{1}{r} \, 
  \ln \left[ \frac{a(t)}{a(0)} \ \frac{r - a(0)}{r - a(t)} \right] \, .
\end{equation*}
Solving for the particle number at time $t$ gives
\begin{equation}
  a(t) = \frac{a(0)}{e^{\left( \mu - \sigma \right) t} 
  \left[ 1 - a(0) / r \right] + a(0) / r} \ .
\label{mfsdp}
\end{equation}
As anticipated, this results in exponential decay for $\mu > \sigma$ with
characteristic time $\tau = 1/| \mu - \sigma | = 1 / \lambda \, |r|$; in the 
active state, the particle number also approaches the carrying capacity $r$
exponentially with rate $1 / \tau$, and monotonically from above or below
for $a(0) > r$ and $a(0) < r$, respectively.
Right at the extinction threshold $\sigma = \mu$ ($r = 0$) separating the
active and inactive and absorbing states, \eref{reqdp} reduces to \eref{reqpa}
for pair annihilation, and hence the exponential kinetics of \eref{mfsdp} is
replaced by the power law decay (\ref{mfspa}), as follows also from taking
the limit $\mu - \sigma \to 0$ in \eref{mfsdp}.

\subsection{Diffusive spreading and spatial clustering}

At least in a phenomenological manner, the above analysis can be readily
generalized to spatially extended systems by adding (e.g., unbiased 
nearest-neighbor) particle hopping or exchange in lattice models, or diffusive 
spreading in a continuum setting.
Still within a mean-field framework that entails mass action factorization of
non-linear correlations, in the continuous representation this leads to a 
\emph{reaction-diffusion equation} 
\begin{equation}
  \frac{\partial a(\vec{x},t)}{\partial t}  \approx \left( \sigma - \mu 
 + D \nabla^2\right) a(\vec{x},t) - {\bar \lambda} \, a(\vec{x},t)^2 
\label{rdedp}
\end{equation}
for the \emph{density field} $a(\vec{x},t)$ with diffusion constant $D > 0$.
In the population dynamics context, this non-linear partial differential 
equation is referred to as the Fisher--Kolmogorov(--Petrovskii--Piskunov)
equation \cite{jdm02}.
In one spatial dimension, \eref{rdedp} admits solitary traveling wave 
solutions of the form $a(x,t) = u(x - c \, t)$ that interpolate between the
active and inactive states, i.e., $u(z \to -\infty) = 0$, while 
$u(z \to \infty) = {\bar r} = (\sigma - \mu) / {\bar \lambda}$, if 
${\bar r} > 0$; their detailed shape depends on the wave velocity $c$.

Away from the extinction threshold at $\sigma = \mu$, we may linearize this
equation by considering the deviation 
$\delta a(\vec{x},t) = a(\vec{x},t) - a(\infty)$ from the asymptotic density
$a(\infty)$.
Upon neglecting quadratic terms in the fluctuations $\delta a$, one obtains
near both the active (where $a(\infty) ={\bar r}$) and inactive states (with 
$a(\infty) = 0$)
\begin{equation}
  \frac{\partial \delta a(\vec{x},t)}{\partial t} \approx \left( D \nabla^2 -
  | \sigma - \mu | \right) a(\vec{x},t) \ ,
\label{lrddp}
\end{equation}
with characteristic length scale 
$\xi = \sqrt{D / | \sigma - \mu |}$ and corresponding time scale 
$\tau = \xi^2 / D$ as to be expected for diffusive processes.
The correlation length $\xi$ describes the extent of spatially correlated 
regions, i.e., density clusters, in the system, whereas the rate $1 / \tau$
governs their temporal decay.

Since $a(t) = \int a(\vec{x},t) \, d^dx$, the mean-field logistic equation
(\ref{reqdp}) follows from \eref{rdedp} only under the assumption of 
extremely short-range spatial correlations $\sim \delta(\vec{x} - \vec{x}')$,
i.e., in the limit $\xi \to 0$, which is definitely violated near the extinction
threshold.
This Dirac delta function also indicates that the non-linear reaction rate in
the continuum description and the corresponding dimensionless mean-field
rate are related to each other via the volume $b^d$ of the unit cell in an 
ultimately underlying discrete lattice model: 
$\lambda \sim b^d \, {\bar \lambda}$.
It is important to realize that the connection between microscopic reaction
rates and their continuum counterparts is not usually direct and simple, but 
depends on the details of the involved coarse-graining process.
With this caveat stated explicitly, which also applies to the relationship 
between lattice hopping rates and continuum diffusivities, we shall for
notational simplicity henceforth drop the overbars from the continuum rates.

\subsection{Extinction threshold: directed percolation criticality}

On a lattice with $N$ sites, where locally spontaneous particle death,
reproduction, and binary coagulation (or annihilation) can take place, and
which are coupled through particle hopping processes, the single-site 
extinction bifurcation discussed above translates into a genuine continuous
non-equilibrium phase transition in the thermodynamic limit $N \to \infty$, 
or in the corresponding continuum model with infinitely many degrees of
freedom.
Note that in any finite stochastic system that incorporates an absorbing state,
the latter is inevitably reached at sufficiently long times.
A true absorbing-to-active phase transition hence requires taking the 
thermodynamic limit first in order to permit the existence of a stable active
phase as $t \to \infty$.
In addition, this asymptotic long-time limit must be considered prior to tuning
any control parameters.
Yet characteristic extinction times tend to grow exponentially with $N$ and
active states may survive in a quasi-stationary configuration as long as
$\log t \ll {\cal O}(N)$.
Hence absorbing phase transitions are in fact easily accessible numerically in 
computer simulations with sufficiently many lattice sites.
As in the vicinity of critical points or second-order phase transitions in thermal
equilibrium, non-linearities and fluctuations become crucial near the extinction
threshold, and cannot be neglected in a proper mathematical treatment.
Consequently, mean-field approximations that neglect both intrinsic reaction
noise and spatial correlations become at least questionable.

The phenomenological description of active-to-absorbing (and other
non-equilibrium) phase transitions closely follows that of near-equilibrium
critical dynamics \cite{hh01, go04, jt05, thv05, hhl08, uct14}.
As the phase transition is approached upon tuning a relevant control 
parameter $r \to 0$, spatial correlations become drastically enhanced.
For the correlation function of an appropriately chosen order parameter field
that characterizes the phase transition (e.g., the particle density in our
population dynamics model), one expects a typically algebraic divergence of
the associated correlation length: $\xi(r) \sim |r|^{- \nu}$ with a critical
exponent $\nu$.
Consequently, microscopic length (and time as well as energy) scales are
rendered irrelevant, and the system asymptotically becomes scale-invariant.
This emergent critical-point symmetry is reflected in power law behavior for
various physical quantities that are captured through additional critical indices.
The \emph{dynamic critical exponent} $z$ links the divergence of the
characteristic relaxation time to that of the correlation length: 
$\tau(r) \sim \xi(r)^z \sim |r|^{- z \nu}$, describing \emph{critical 
slowing-down}.
The stationary (long-time) order parameter sets in algebraically: 
$a(t \to \infty, r) \sim r^\beta$ for $r > 0$, while it decays to zero as 
$a(t, r = 0) \sim t^{- \alpha}$ precisely at the critical point.
These two power laws are limiting cases of a more general \emph{dynamical
scaling} ansatz for the time-dependent order parameter,
\begin{equation}
  a(t,r) = |r|^\beta \, {\hat a}\bigl( t / \tau(r) \bigr) \ ,
\label{dynsc}
\end{equation}
where the scaling function on the right-hand side satisfies ${\hat a}(0) =$ 
const.
For large arguments $y = t / \tau(r) \sim t \, |r|^{z \nu}$, one must require 
${\hat a}(y) \sim y^{- \beta / z \nu}$ in order for the $r$ dependence to
cancel as $r \to 0$.
Hence we obtain the critical decay exponent $\alpha = \beta / z \nu$.
This \emph{scaling relation} is of course fulfilled by the mean-field critical
exponents $\alpha = \beta = 1$,  $\nu = 1/2$, and $z = 2$ (indicating
diffusive spreading) found in our previous population model analysis.

As for equilibrium critical phenomena, there exists an (upper) \emph{critical
dimension} $d_c$ below which fluctuations are strong enough to modify not
just amplitudes and scaling functions, but alter the critical scaling exponents.
We invoke a simple scaling argumens to determine $d_c$ for competing
birth-death-coagulation processes: 
Let us set an inverse length (wave vector) scale $\kappa$, i.e., 
$[x] = \kappa^{-1}$, and corresponding time scale 
$[t] = [x]^2 = \kappa^{-2}$ (implying that we choose $[D] = \kappa^0$),
where the square bracket indicates the scaling dimension.
\eref{rdedp} then enforces $[\sigma] = [\mu] = [r] = \kappa^2$; moreover,
$[a(\vec{x},t)] = \kappa^d$ since $a$ represents a density field in $d$
spatial dimensions, whence we find $[\lambda] = \kappa^{2 - d}$ for the
coagulation (or annihilation) rate.
Non-linear stochastic fluctuations that will affect particle propagation and
(linear) extinction incorporate subsequent branching and coagulation
processes, and hence scale like the rate product 
$[\sigma \, \lambda] = \kappa^{4 - d}$.
The critical dimension $d_c = 4$ indicates when this effective non-linear
coupling becomes scale-invariant.
For $d < d_c$, it attains a positive scaling dimension and is considered 
\emph{relevant} (in the renormalization group sense), along with the
`mass'-like parameter $r$.
In dimensions beyond $d_c$, the non-linearity becomes irrelevant and does
not alter the fundamental scaling properties of the model, yet of course 
fluctuations still contribute numerically to various observables.
At $d = d_c$, one usually finds \emph{logarithmic corrections} to the 
mean-field power laws \cite{hhl08, uct14}.

In order to heuristically include fluctuations, one may add (for simplicity)
Gaussian white noise with vanishing average 
$\langle \zeta(\vec{x},t) \rangle = 0$ to the reaction-diffusion equation
(\ref{rdedp}), turning it into a stochastic partial differential equation
\cite{jt05, uct14}:
\begin{equation}
  \frac{\partial a(\vec{x},t)}{\partial t} = \left( \lambda \, r + D \nabla^2
  \right) a(\vec{x},t) - \lambda \, a(\vec{x},t)^2 + \zeta(\vec{x},t) \ .
\label{landp}
\end{equation}
Note that the deterministic part of the right-hand side can be interpreted as an
expansion of a very general reaction functional in terms of the small
fluctuating local particle density near the extinction threshold, and even the
diffusive spreading term may be viewed as the leading contribution in a 
long-wavelength expansion for spatially varying fluctuations (in systems with
spatial inversion symmetry).
\eref{landp} is thus quite generic, provided there are no additional special
symmetries that would enforce the coefficients $r$ or $\lambda$ to vanish.
For $r > 0$ and if $\lambda > 0$, the system resides in an active phase, 
whereas it reaches the empty, absorbing state for $r < 0$.
For negative $\lambda$, one would have to amend \eref{landp} with a cubic
term in the density field.
The absence of a constant particle source (or sink) term on its right-hand side 
is mandated by presence of the absorbing state $a = 0$.

\begin{figure}[ht]
\centerline{\includegraphics[width = 6cm]{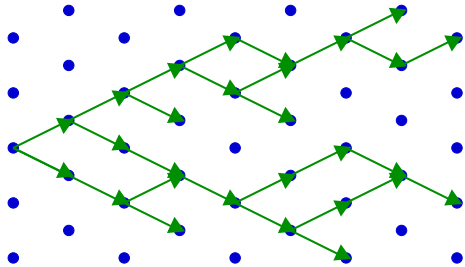}}
\caption{Elementary death, birth, hopping, and coagulation processes
  initiated by a single particle seed generate a directed percolation cluster;
  bonds connecting lattice sites are formed only by advancing along the 
  `forward' (time-like) direction to the right.
  [Figure reproduced with permission from H.~K. Janssen and 
   U.~C. T\"auber, \emph{Ann. Phys. (NY)} {\bf 315}, 147--192 (2005); 
   DOI: 10.1016/j.aop.2004.09.011; copyright (2005) by Elsevier Inc.]}  
\label{dpfig}
\end{figure}

This constraint must be similarly reflected in the noise correlations: as the
mean particle number $a(t) \to 0$, all stochastic fluctuations must cease.
To lowest non-vanishing order in $a$, one would therefore posit 
\begin{equation}
  \left\langle \zeta(\vec{x},t) \, \zeta(\vec{x}',t') \right\rangle = 
  v \, a(\vec{x},t) \, \delta(\vec{x}-\vec{x}') \delta(t-t') \ .
\label{noidp}
\end{equation}
Stochastic dynamics with such a multiplicative noise correlator is properly
defined through a corresponding functional integral representation \cite{jt05,
uct14}; intriguingly, the ensuing path integral action turns out to be 
equivalent to a well-studied problem in nuclear and particle physics, namely 
Reggeon field theory \cite{mm78}, which in turn has been shown to capture 
the universal scaling properties of critical \emph{directed percolation} 
clusters \cite{cs80, hkj81}.
This sequence of mathematical mappings lends strong support to the 
Janssen--Grassberger conjecture \cite{hkj81, pg82}, which states that the
asymptotic critical scaling features of continuous non-equilibrium transitions
from active to inactive, absorbing states for a single scalar order parameter
governed by Markovian stochastic dynamics, and in the absence of any
quenched disorder and coupling to other conserved fields, should be
described by the directed-percolation universality class.
In fact, this statement applies generically even for multi-component systems
\cite{hkj01}.
Indeed, if we represent the basic stochastic death, birth, nearest-neighbor
hopping, and coagulation processes on a lattice through their `world lines' in
a space-time plot as depicted in Fig.~\ref{dpfig}, it becomes apparent how
they generate a directed percolation cluster \cite{jt05, hhl08}.

More rigorously, the Doi--Peliti formalism allows a representation of 
`microscopic' stochastic reaction kinetics as defined through the associated
master equation in terms of a coherent-state path integral \cite{thv05, uct14}.
For the combined reactions $A \to \emptyset$, $A \to A + A$, and 
$A +A \to A$ in the continuum limit, augmented with diffusive spreading,
the resulting action assumes precisely the form of Reggeon field theory, and
hence the `mesoscopic' Langevin description (\ref{landp}, \ref{noidp}), in
the vicinity of the absorbing-state transition \cite{jt05, thv05, uct14}.
As a consequence of a special internal `rapidity reversal' symmetry of 
Reggeon field theory, the directed-percolation universality class is fully
characterized by a set of only three independent critical exponents (say $\nu$,
$z$, and $\beta$); all other critical indices are related to them via exact
scaling relations.

The critical exponents for directed percolation are known to high precision
from dedicated Monte Carlo computer simulations \cite{hh01, go04, hhl08};
recent literature values for $\alpha, \beta, \nu$, and $z$ in two and three
dimensions are listed in Table~\ref{dpexp}.
For $d = 1$, the most accurate exponent values were actually obtained 
semi-analytically via ingenious series expansions \cite{ij96}.
A dynamical renormalization group analysis based on the (Reggeon) field
theory representation allows a systematic perturbation expansion for the
fluctuation corrections to the mean-field exponents in terms of the deviation 
$\epsilon = 4 - d$ from the critical dimension $d_c$ \cite{mm78, jt05,
thv05, uct14}.
The first-order (one-loop) results are also tabulated below; as compared to 
mean-field theory, critical fluctuations effectively reduce the values of 
$\alpha, \beta, 1/\nu$, and $z$ in accord with the numerical data, and
increasingly so for lower dimensions.
Note that $z < 2$ implies sub-diffusive propagation at the extinction
transition for $d < 4$.
Several experiments have at least partially detected directed-percolation
scaling near active-to-absorbing phase transitions in various systems
\cite{hhl08}.
Perhaps the most impressive confirmation originates from detailed and very
careful studies of the transition between two different turbulent states of
electrohydrodynamic convection in (quasi-)two-dimensional turbulent nematic
liquid crystals carried out by Takeuchi et al. at the University of Tokyo
\cite{tkcs09}, who managed to extract twelve different critical exponents (four
of them listed in Table~\ref{dpexp}) along with five scaling functions from
their experimental data.

\begin{table}[ht]
\tbl{Critical exponents for directed percolation obtained from series
  expansions in one dimension \cite{ij96}; Monte Carlo simulations in two
  and three dimensions \cite{hhl08}; first-order perturbative renormalization
  group analysis ($d = 4 - \epsilon$) \cite{jt05}; and experiments on   
  turbulent nematic liquid crystals ($d = 2$) \cite{tkcs09} (numbers in
  brackets indicate last-digit uncertainties).} 
  {\begin{tabular}{cccccc} \toprule 
  critical & & dimension & & & liquid crystal \\ exponent & $d = 1$ & 
  $d = 2$ & $d = 3$ & $d = 4 - \epsilon$ & experiment \\
  \colrule $\alpha$ & $0.159464(6)$ & $0.4505(10)$ & $0.732(4)$ & 
  $1 - \frac{\epsilon}{4 }+ {\cal O}(\epsilon^2)$ & 0.48(5) \\ 
  $\beta$ & $0.276486(8)$ & $0.5834(30)$ & $0.813(9)$ & 
  $1 - \frac{\epsilon}{6} + {\cal O}(\epsilon^2)$ & 0.59(4) \\ 
  $\nu$ & $1.096854(4)$ & $0.7333(75)$ & $0.584(5)$ & 
  $\frac12 + \frac{\epsilon}{16} + {\cal O}(\epsilon^2)$ & 0.75(6) \\
  $z$ & $1.580745(10)$ & $1.7660(16)$ & $1.901(5)$ & 
  $2 - \frac{\epsilon}{12} + {\cal O}(\epsilon^2)$ & 1.72(11) \\ \botrule
  \end{tabular}}
\label{dpexp}
\end{table}

\section{Dynamic Correlations in Pair Annihilation Processes}

\subsection{Depletion zones and reaction rate renormalization}

We next return to simple pair annihilation ($A + A \to \emptyset$, rate 
$\lambda'$) or fusion ($A + A \to A$, rate $\lambda$) processes, but in
spatially extended systems.
The on-site master equation for the former looks like \eref{masan}, with the
gain term replaced with $\lambda' \, (n+1) (n+2) \, P(n+2)$.
This merely results in a rescaled reaction rate $\lambda \to 2 \lambda'$ in
both the exact \eref{exqpa} and the mean-field equations (\ref{reqpa}),
(\ref{mfspa}).
However, as the particle density drops toward substantial dilution, the rate of
further pair annihilation processes will ultimately not be determined by the
original microscopic reactivity $\lambda$ ($\lambda'$), and instead be 
limited by the time it takes for two reactants to meet \cite{kk88, otb89,
krb10, thv05}.
If we assume diffusive spreading with diffusion constant $D$, and hence
relative diffusivity $2 D$, the typical time for two particles at distance $l$ to
find each other is $t \sim l^2 / 4 D$.
In the diffusion-limited regime, $l$ and $t$ set the relevant length and time
scales, whence the particle density should scale according to 
$a(t) \sim l(t)^{-d} \sim (D \, t)^{- d / 2}$ in $d$ spatial dimensions.
This suggests a slower decay than the mean-field prediction 
$a(t) \sim (\lambda \, t)^{-1}$ in dimensions $d < 2$.
Indeed, the previously established scaling dimension 
$[\lambda] = \kappa^{2 - d}$ too indicates that $d_c = 2$ sets the critical
dimension for diffusion-controlled pair annilation.
In one dimension, the certain return of a random walker to its origin ensures
that the ensuing annihilations carve out \emph{depletion zones}, generating
spatial \emph{anti-correlations} that impede subsequent reactions.
In contrast, for $d > 2$ diffusive spreading sustains a well-mixed system with
largely homogeneous density, and mean-field theory remains valid.

This simple scaling argument is quantitatively borne out by Smoluchowski's
classical self-consistent approach \cite{thv05, krb10}.
In a continuum representation, we need to impose a finite reaction sphere
with radius $b$: two particles react, once their distance becomes smaller than
$b$.
In a quasi-stationary limit, we then need to solve the stationary diffusion 
equation
\begin{equation}
  0 = \nabla^2 a(r) = 
  \frac{\partial^2 a(r)}{\partial r^2} + 
  \frac{d - 1}{r} \, \frac{\partial a(r)}{\partial r} \ ,
\label{stdif}
\end{equation}
where we have invoked spherical symmetry and written down the Laplacian 
differential operator in $d$-dimensional spherical coordinates.
General solutions are then linear combinations of a constant term and the
power $r^{2 - d}$.
For $d > 2$, we may impose the straightforward boundary conditions
$a(r \leq b) = 0$, whereas the particle density approaches a finite asymptotic 
value $a(\infty)$ far away from the reaction center located at the origin, which
yields $a(r) = a(\infty) \left[ 1 - (b / r)^{d - 2} \right]$.
The effective reactivity ${\widetilde \lambda}$ in the diffusion-limited regime
is then given by the incoming particle flux at reaction sphere boundary:
${\widetilde \lambda} \sim D b^{d - 1} a(\infty)^{-1} 
 \left[ \partial a(r) /  \partial r \right] |_{r = b} \sim D (d - 2) b^{d - 2}$,  
which replaces the annihilation rate $\lambda$ in the rate equation 
(\ref{reqpa}). 
Consequently, one obtains the large-time density decay 
$a(t) \sim (D \, t)^{-1}$ with the same power law (\ref{mfspa}) as in the 
reaction-controlled region at large densities.

However, at $d_c = 2$ the effective reaction rate ${\widetilde \lambda}$
vanishes; indeed, in low dimensions $d < 2$ one needs to impose a different
boundary condition $a(R) \approx a(\infty)$, where $R$ denotes the mean
particle separation: since the density $a(R) \sim R^{-d}$, one has 
$R(a) \sim a^{- 1 / d}$.
The boundary condition hence depends on the actual density in the 
quasi-stationary limit, resulting in the profile
$a(r) = a(\infty) \left[ (r / b)^{2 - d} - 1 \right] / \left[ (R / b)^{2 - d} - 1   
 \right]$.
From the diffusive flux at the reaction sphere one obtains the effective 
reactivity 
${\widetilde \lambda}(a) \sim D (2 - d) b^{d - 2} / [(R / b)^{2 - d} - 1] \to 
 \lambda_R \,  a^{-1 + 2 / d}$ as $a \to 0$ and $R \to \infty$, with a 
constant $\lambda_R \sim D$ (that again tends to zero as $d \to 2$).
Upon self-consistently replacing $\lambda$ with the density-dependent
effective rate ${\widetilde \lambda}(a)$ in \eref{reqpa}, we arrive at
\begin{equation}
  \frac{\partial a(t)}{\partial t} \sim - \lambda_R \, a(t)^{1 + 2 / d} \ .
\label{reqsm}
\end{equation}
It solution through variable separation yields the anticipated slower density
decay and in turn the time dependence of the effective reaction rate:
\begin{equation}
  a(t) \sim (D \, t)^{- d / 2} \ , \quad 
  {\widetilde \lambda}(t) \sim (D \, t)^{-1 + d/2} \ .
\label{solsm}
\end{equation}
At the critical dimension $d_c = 2$, one finds logarithmic slowing-down
relative to the rate equation power law:  
$a(t) \sim (D \, t)^{-1} \ln (8 D \, t / b^2)$.

Dynamical renormalization group calculations based on the Doi--Peliti field
theory representation of the associated master equation confirm these 
findings \cite{lp86, bpl94, thv05, uct14}.
For $k$th order annihilation $k \, A \to \emptyset, A, \ldots, (k - 1) A$ 
($k \geq 2$) the right-hand side of \eref{reqpa} is to be replaced with 
$- \lambda_k \, a(t)^k$; for diffusive propagation, dimensional analysis then
yields the scaling dimension $[\lambda_k] = \kappa^{2 - (k - 1) \, d}$, from
which one infers the upper critical dimension $d_c(k) = 2 / (k - 1)$.
Deviations from the mean-field algebraic decay $a(t) \sim t^{- 1 / (k - 1)}$
should only materialize for pair ($k = 2$) reactions for $d \leq 2$, and for
triplet annihilationat $d_c(3) = 1$.
In accord with Smoluchowski's underlying assumption, diffusive spreading is
not affected by the non-linear annihilation processes.
The resulting perturbation expansion in $\lambda_k$ can be summed to all
orders, and indeed recovers \eref{solsm} for pair annihilation, and the 
corresponding logarithmic scaling at the critical dimension, which for triplet
reactions becomes $a(t) \sim \left[ (D \, t)^{-1} \ln (D\, t) \right]^{1/2}$.

Pair annihilation dynamics on one-dimensional lattices with strict site
exclusion can also be mapped onto non-Hermitean spin-$1/2$ Heisenberg
models, permitting the extraction of remarkably rich non-trivial exact results 
\cite{adhr94, hos97, gms01, rs01}.   
The anomalous density decay induced by the self-generated depletion zones 
has been confirmed in several experiments on exciton recombination kinetics
in effectively one-dimensional molecular systems.
Particularly convincing are data obtained by Allam et al. at the University of
Surrey in carbon nanotubes, who managed to explore the detailed crossover
in the power laws from the reaction-controlled to the diffusion-limited regime
(\ref{solsm}) both in the exciton density decay and the reactivity \cite{ja13}.

\subsection{Segregation in diffusive two-species annihilation}

Let us now investigate pair annihilation of particles of distinct species,
$A + B \to \emptyset$, with reaction rate $\lambda$ \cite{hh83, jdm02,
krb10, uct14}.
The crucial distinction to the previous situation is that alike particles do not
react with each other. 
The associated exact time evolution as well as the coupled mean-field rate
equations for the densities $a(t)$ and $b(t)$ are symmetric under species
exchange $A \leftrightarrow B$,
\begin{equation}
  \frac{\partial a(t)}{\partial t} = \frac{\partial b(t)}{\partial t} =
  - \lambda \left\langle [n_A \, n_B](t) \right\rangle \approx 
  - \lambda \, a(t) \, b(t) \ , 
\label{reqab}
\end{equation}
and these binary reactions of course preserve the number difference
$n_C = n_A - n_B =$ const., whence also $c(t) = a(t) - b(t) = c(0)$.
One must now distinguish between two situations:
If initially $n_A = n_B$ precisely, $c(t) = 0$ at all times:
with identical initial conditions for their same rate equations, $a(t) = b(t)$.
\eref{reqab} consequently reduces to \eref{reqpa} for pair annihilation of
identical species, with the solution (\ref{mfspa}) that describes algebraic
decay to zero for both populations.
In the more generic case $n_A \not= n_B$, say, with majority species $A$,
i.e, $c(0) > 0$, the $B$ population will asymptotically go extinct, 
$b(\infty) = 0$, while $a(\infty) = c(0)$.
At long times, we may thus replace $a(t) \approx c(0)$ in the rate equation
for $b(t)$, resulting in exponential decay with time constant $c(0) \lambda$:
$b(t) = a(t) - c(0) \sim e^{- c(0) \lambda \, t}$.
The special symmetric case $c(0) = 0$ hence resembles a dynamical 
critical point with diverging relaxation time, and exponential density decay
replaced by a power law.

In a spatial setting with diffusive transport, ultimately the stochastic pair
annihilation reactions will become diffusion-limited, and the emerging 
depletion zones and persistent particle anti-correlations will markedly slow 
down the asymptotic decay of the minority species in low dimensions 
$d \leq d_c = 2$.
Indeed, replacing 
$\lambda \, t \to {\widetilde \lambda}(t) \, t \sim (D \, t)^{d / 2}$ with its 
renormalized counterpart according to \eref{solsm}, one arrives at stretched
exponential behavior: $\ln b(t) = \ln [a(t) - c(0)] \sim - (D \, t)^{d / 2}$ for
$d < 2$,  whereas $\ln b(t) = \ln [a(t) - c(0)] \sim - (D \, t) / \ln (D \, t)$ in
two dimensions. 

In the special symmetric case with equal initial particle numbers, the presence
of an additional conserved quantity has a profound effect on the long-time
chemical kinetics: 
Both particle species may spatially \emph{segregate} into inert, slowly
coarsening domains, whence the reactions become confined to the contact 
zones separating the $A$- or $B$-rich domains \cite{tw83, lc95}.
Note that the local particle density excess satisfies a simple diffusion equation
$\partial c(\vec{x},t) / \partial t = D \nabla^2 c(\vec{x},t)$; the associated
initial value problem is then solved by means of the diffusive Green's function
$G(\vec{x},t) = \Theta(t) \, e^{- \vec{x}^2 / 4 D \, t} / 4 \pi D \, t$ via the
convolution integral
$c(\vec{x},t) = \int G(\vec{x} - \vec{x}',t) \, c(\vec{x}',0) \, d^dx'$.
If one assumes an initially random, spatially uncorrelated Poisson distribution
for both $A$ and $B$ particles with 
$\overline{a(\vec{x},0)} = \overline{b(\vec{x},0)}  = a(0)$,
where the overbar denotes an ensemble average over initial conditions, and
$\overline{a(\vec{x},0) \, a(\vec{x}',0)} = \overline{b(\vec{x},0) \,
  b(\vec{x}',0)} = a(0)^2 + a(0) \, \delta(\vec{x} - \vec{x}')$, whereas 
$\overline{a(\vec{x},0) \, b(\vec{x}',0)} = 0$, the corresponding moments
for the initial density excess become $\overline{c(\vec{x},0)} = 0$ and
$\overline{c(\vec{x},0) \, c(\vec{x}',0)} = 
 2 a(0) \, \delta(\vec{x} - \vec{x}')$. 

These considerations allow us to explicitly evaluate 
\begin{equation}
  \overline{c(\vec{x},t)^2} = 2 \, a(0) \int G(\vec{x} - \vec{x}'t)^2 \, d^dx'
  = \frac{2 \, a(0) \, \Theta(t)}{\left( 8 \pi D \, t \right)^{d / 2}}
\label{abvar}
\end{equation} 
through straightforward Gaussian integration.
The distribution of the field $c$ itself will be Gaussian as well, with zero mean
and variance (\ref{abvar}); hence we finally obtain for the average absolute
value of the local density excess
\begin{equation}
  \overline{|c(\vec{x},t)|} = \sqrt{\frac{2}{\pi} \ \overline{c(\vec{x},t)^2}}
 = \sqrt{\frac{4 \, a(0)}{\pi}} \ 
 \frac{\Theta(t)}{\left( 8 \pi D \, t \right)^{d / 4}} \ .
\end{equation}
In high dimensions $d > 4$, this excess decays faster than the mean 
densities  $a(t) \sim 1 / t$, implying that the particle distribution remains
largely uniform, and mean-field theory provides a satisfactory description.
In contrast, for $d < d_s = 4$, the long-time behavior is dictated by the
slowly decaying spatial density excess fluctuations:
$a(t) \sim b(t) \sim (D \, t)^{- d / 4}$.
Either species accumulate in diffusively growing domains of linear size 
$l(t) \sim (D \, t)^{1/2}$ separated by active \emph{reaction zones} whose
width relative to $l(t)$ decreases algebraically with time \cite{lc94}. 
Note that the borderline dimension for spatial species segregation $d_s = 4$
is not a critical dimension in the renormalization group sense; the anomalous
density decay as consequence of segregated domain formation can rather be
fully described within the framework of mean-field reaction-diffusion 
equations \cite{lc95}.
The resulting power law $a(t) \sim t^{- 3 / 4}$ in three dimensions was
experimentally observed in a calcium-fluorophore system by Monson and
Kopelman at the University of Michigan \cite{mk04}.

The above analysis does not apply to specific, spatially correlated initial
conditions.
For example, if impenetrable hard-core particles are aligned in strictly
alternating order $\cdots A B A B A B A \cdots$ on a one-dimensional line,
pair annihilation reactions will preserve this arrangement at all later times.
Hence the distinction between the two species becomes in fact meaningless,
and the single-species asymptotic decay $\sim (D \, t)^{- 1 / 2}$ ensues.
More generally, pair annilation processes involving $q$ distinct species 
$A_i + A_j \to \emptyset$ ($1 \leq i < j \leq q$) should eventually reduce to
just a two-species system for the remaining two `strongest' particle types, as
determined by their reaction rates, diffusivities, and initial concentrations.
Novel, distinct behavior could thus only appear for highly symmetric
situations where all reaction and diffusion rates as well as initial densities are
set equal among the $q$ species.

Yet it turns out that for any $q \geq 3$ there exist no conserved quantities in
such systems.
One may furthermore establish the borderline dimension for species 
segregation as $d_s(q) = 4 / (q - 1)$; for $d \geq 2$, therefore all densities
should follow the mean-field decay law $\sim 1 / t$ as for a single species
\cite{hdwt04}.  
Only in one dimension can distinct particle species cluster into stable 
domains, resulting in a combination of depletion- and segregation-dominated
decay:
\begin{equation}
  a_i(t) \sim t^{- 1 / 2} + C \, t^{- \alpha(q)} \ , \quad 
  \alpha(q) = \frac{q - 1}{2 \, q} \ .
\label{qsppa}
\end{equation}
Note that $\alpha(2) = 1 / 4$ as established above, while the single-species
decay exponent is recovered in the limit of infinitely many species, 
$\alpha(\infty) = 1/2$: 
In that situation, alike particles experience a vanishing probability to ever 
encounter each other, whence the distinction between the different species 
becomes irrelevant.
Once again, correlated initial linear arrangements such as 
$\cdots A B C D A B C D \cdots$ for four species induce special cases; here,
no alike particles can ever meet, and the system behaves effectively like 
single-species pair annihilation again.
Also, interesting cyclic variants may be constructed \cite{hwt04}, such as
$A + B \to \emptyset$, $B + C \to \emptyset$, $C + D \to \emptyset$, and
$D + A \to \emptyset$.
In this case, one may collect individuals from species $A$ and $C$, and
similarly $B$ and $D$, in just two competing `alliances', and consequently 
the decay kinetics is captured by the two-species pair annihilation behavior. 

\section{Stochastic Pattern Formation in Population Dynamics}

\subsection{Activity fronts in predator-prey coexistence models}

The language and also numerical and mathematical tools developed for the
investigation of chemical kinetics may be directly transferred to spatially
extended stochastic population dynamics \cite{mgt07, uct14, dmpt18}.
Aside from the logistic population growth with finite carrying capacity 
(\ref{reqdp}), another classical and paradigmatic model in ecology concerns
the competition and coexistence of prey with their predator species, as first
independently constructed by Lotka and Volterra \cite{hh83, jdm02}.
Let $B$ indicate the prey species, who left on their own merely undergo
(asexual) reproduction $B \to B + B$ with rate $\sigma$.
Their population is held in check by predators $A$ who may either 
spontaneously die, $A \to \emptyset$ with rate $\mu$, or upon encounter
with a prey individual, devoure it and simultaneously generate offspring:
$A + B \to A + A$ with rate $\lambda$.
The original deterministic Lotka--Volterra model consists of the associated
coupled mean-field rate equations for the population densities $a(t)$ and
$b(t)$:
\begin{equation}
  \frac{\partial a(t)}{\partial t} = - \mu \, a(t) + \lambda \, a(t) \, b(t) \ , 
  \quad 
  \frac{\partial b(t)}{\partial t} = \sigma \, b(t) - \lambda \, a(t) \, b(t) \ .
\label{reqlv}
\end{equation}
This dynamics allows three stationary states, namely (i) complete extinction 
$a = b = 0$; (ii) the also absorbing pure prey state with Malthusian 
population explosion $a = 0$, $b \to \infty$; and (iii) a predator-prey 
coexistence state with finite densities $a(\infty) = \sigma / \lambda$,
$b(\infty) = \mu / \lambda$.
Naturally, the predators benefit from high prey fertility $\sigma$, while the 
prey prosper if the predators are short-lived; yet counter-intuitively for the
predators, both stationary population numbers decrease with enhanced 
predation rates $\lambda$, signaling a non-linear feedback mechanism:
If the $A$ species too efficiently reduces the $B$ population, they have scarce
food left, whence the majority of them die.

However, this stationary coexistence state (iii) is in fact never reached under 
the deterministic non-linear dynamics (\ref{reqlv}).
Indeed, eliminating time through taking the ratio $da / db = 
 \left( \lambda \, b - \mu \right) a / \left( \sigma - \lambda \, a \right) b$,
one obtains after variable separation and integration a \emph{conserved first
integral} for the mean-field dynamics: $K(t) = 
 \lambda \left[ a(t) + b(t) \right] - \sigma \ln a(t) - \mu \ln b(t) = K(0)$. 
The trajectories in the phase space spanned by the population numbers must
therefore be strictly periodic orbits, implying undamped non-linear population
oscillations whose amplitudes and shapes are fixed by the initial values $a(0)$
and $b(0)$.
For small deviations from the stationary coexistence center 
$\delta a(t) = a(t) - a(\infty)$, $\delta b(t) = b(t) - b(\infty)$, straightforward
linearization of Eqs.~(\ref{reqlv}) yields
$\partial \delta a(t) / \partial t \approx \sigma \, \delta b(t)$,
$\partial \delta b(t) / \partial t \approx - \mu \, \delta a(t)$, which are then
readily combined to the simple harmonic oscillator differential equation
$\partial^2 \delta a(t) / \partial t^2 \approx - \omega^2 \, a(t)$ with (linear)
oscillation frequency $\omega = \sqrt{\mu \, \sigma}$.
Equivalently, we may construct the linear stability matrix ${\bf L}$ that
governs the temporal evolution of the fluctuation vector 
${\bf v} = \left( \delta a \ \delta b \right)^T$: 
$\partial {\bf v}(t) / \partial t \approx {\bf L} \, {\bf v}(t)$, where
${\bf L} = \begin{pmatrix} \ 0 & \ \sigma \\ - \mu & \ 0 \end{pmatrix}$
with imaginary eigenvalues $\pm i \omega$.

Clearly the absence of any real part in the stability matrix eigenvalues
represents a degenerate, a-typical situation that should not be robust against
even minor modifications of the model \cite{jdm02}.
For example, in order to render the Lotka--Volterra description more realistic
and prevent any population divergence, one could impose a finite carrying
capacity $r$ for species $B$; at the mean-field level, this alters the second
differential equation in (\ref{reqlv}) to
\begin{equation}
  \frac{\partial b(t)}{\partial t} = \sigma \, b(t) \left[ 1 - b(t) / r \right] 
  - \lambda \, a(t) \, b(t) \ ,
\label{reclv}
\end{equation}
leading to modified stationary states (ii') $a(\infty) = 0$, $b(\infty) = r$ and
(iii') $a(\infty) = \sigma \left( 1 - \mu / \lambda \, r \right) / \lambda$,
$b(\infty) = \mu / \lambda$.
The latter two-species coexistence fixed point exist, and is linearly stable,
provided the predation rate exceeds the threshold $\lambda_c = \mu / r$;
for $\lambda < \lambda_c$, the predator species $A$ is driven to extinction.
At the stationary state (iii'), the linear stability matrix eigenvalues acquire
negative real parts:
\begin{equation}
  \epsilon_\pm = - \frac{\mu \, \sigma}{2 \, \lambda \, r} \left[ 1 \pm 
  \sqrt{1 - \frac{4 \, \lambda \, r}{\sigma} \left( \frac{\lambda \, r}{\mu} - 1   
  \right)} \, \right] \, , 
\label{coxev}
\end{equation}
For $\sigma > 4 \, \lambda \, r \left( \lambda \, r / \mu - 1 \right)$, these
eigenvalues are both real, indicating exponential relaxation towards the stable
node (iii'). 
For lower prey birth rates, trajectories in phase space spiral inwards to reach
the stationary state (iii') which now represents a stable focus, and the imaginary part of $\epsilon_\pm$ gives the frequency of the resulting damped
population oscillations.
One may amend this mean-field description to allow for spatial structures by
replacing the population numbers with local density fields and adding diffusion terms as in \eref{rdedp}.
In one dimension, the ensuing coupled set of partial differential equations 
permits traveling wave solutions which describe predator invasion fronts 
originating in a region set in the coexistence state (iii') and moving into space
occupied only by prey \cite{jdm02}.

Monte Carlo computer simulations with sufficiently large populations display
rich dynamical features in the (quasi-)stable predator-prey coexistence regime 
that reflect the mean-field picture only partially \cite{mgt07, wmt07, dmpt18}.
Already in a local `urn' model without spatial degrees of freedom, intrinsic
fluctuations, often termed demographic noise in this context, dominate the
dynamics and induce persistent stochastic population oscillations.
These can be understood through the effect of white-noise driving on a 
damped oscillator: 
On occasion, the stochastic forcing will hit the oscillator's eigenfrequency, and
via this resonant amplification displace the phase space trajectories away from
the stable coexistence fixed point \cite{kn05}.
\begin{figure}[ht]
\centerline{
\subfigure[]{\includegraphics[width = 4cm, height = 4cm]{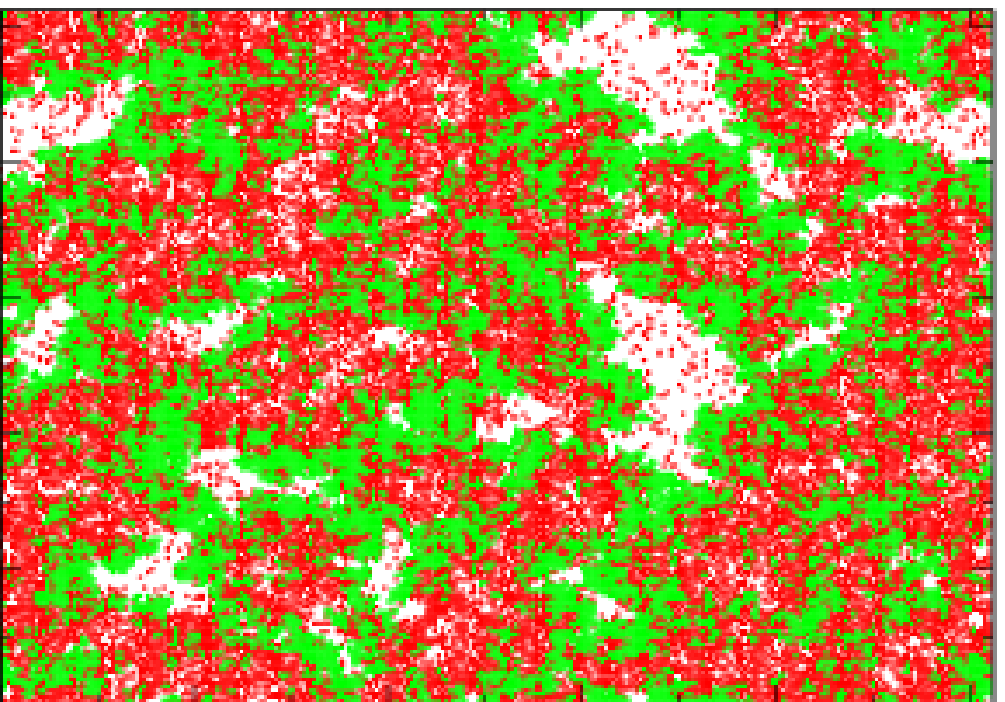}}
\label{lvfiga} \qquad
\subfigure[]{\includegraphics[width = 4cm, height = 4cm]{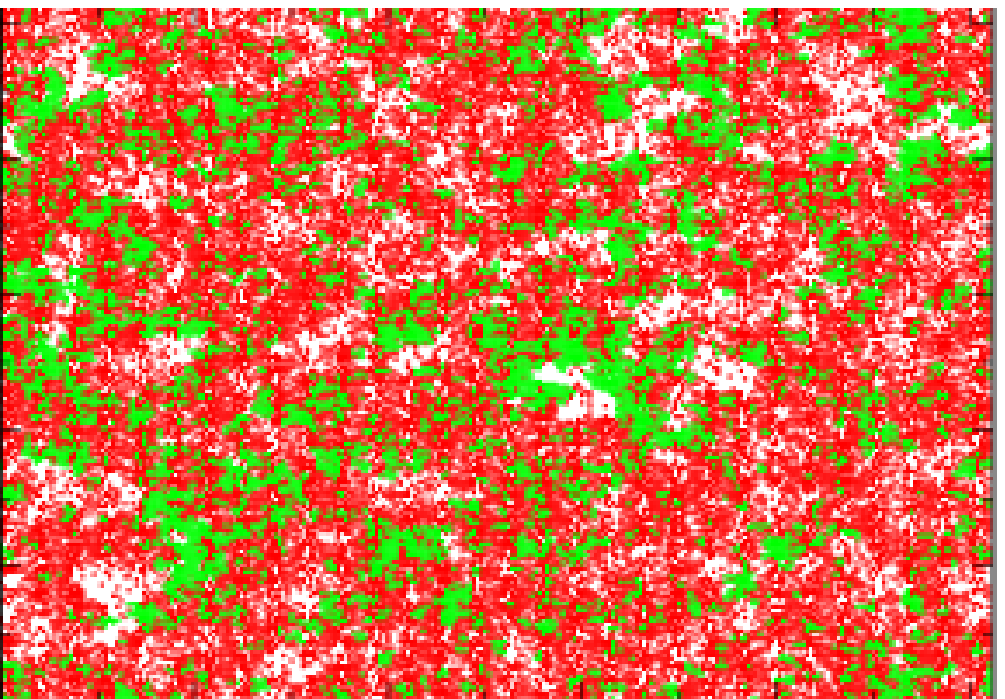}}
\label{lvfigb}}
\caption{Snapshots of a stochastic Lotka--Volterra model on a square lattice
  with $256 \times 256$ sites, periodic boundary conditions, site occupation
  number exclusion, and random initial particle placement after (a) $500$;
  (b) $1000$ Monte Carlo steps.
  Sites occupied by predators are color-coded in red, by prey in green, and
  empty spaces in white.
  [Figures reproduced with permission from S. Chen and U.~C. T\"auber,
   \emph{Phys. Biol.} {\bf 13}, 025005-1--11 (2016);
   DOI: 10.1088/1478-3975/13/2/025005; copyright (2016) by IOP Publ.]}  
\label{lvfig}
\end{figure}

In stochastic Lotka--Volterra models on a lattice that permit an arbitrary
number of particles per site, nearest-neighbor hopping transport generates
wave-like propagation of these erratic local population oscillations 
\cite{wmt07}.
As depicted in the two-dimensional simulation snapshots of Fig.~\ref{lvfig},
prey may thus invade empty regions, followed by predators who feed on
them, and in turn re-generate space devoid of particles in their wake.
Surviving prey islands then act as randomly placed sources of new activity
fronts; it is hence the very stochastic nature of the kinetics that both causes 
and stabilizes these intriguing spatio-temporal patterns \cite{bg09, bg11}.
They are moreover quite robust with respect to modification of the 
microscopic model implementations, and are invariably observed in the
predator-prey coexistence phase in two and three dimensions, even when 
the Lotka--Volterra process $A + B \to A + A$ is separated into independent
predation and predator birth reactions \cite{mgt06}, or if at most a single 
particle is allowed on each lattice site.
For large predation rates, one observes very distinct and prominent activity
fronts; for smaller values of $\lambda$, closer to the extinction threshold,
instead there appear confined fluctuating predator clusters immersed in a sea
of abundant prey.
These two different types of structures may be related to the distinct 
properties of the mean-field linear stability matrix:
Real eigenvalues $\epsilon_\pm$ that indicate purely relaxational kinetics 
correspond to isolated prey clusters in the spatial system, whereas imaginary
parts in \eref{coxev} associated with spiraling trajectories and damped
oscillations pertain to spreading predator-prey fronts.

In either case, analysis of the Fourier-transformed average population 
densities quantitatively establishes that stochastic fluctuations both drastically
renormalize the oscillation frequencies relative to the rate equation prediction
and generate attenuation \cite{mgt07, wmt07}.
Based on the Doi--Peliti mapping of the corresponding master equation to a
continuum field theory \cite{thv05, uct14}, a perturbative computation in 
terms of the predation rate $\lambda$ qualitatively confirms these numerical
observations:
It demonstrates the emergence of a noise-induced effective damping, as well 
as the fluctuation-driven instability towards spatially inhomogeneous 
structures in dimensions $d \leq 4$; the calculation also explains the strong
downward renormalization for the population oscillation frequency as caused by an almost massless mode \cite{uct12}.

Implementing a finite local carrying capacity through constraining the site
lattice occupations, say, to at most a single particle of either type, in addition
produces an extinction threshold for the predator species as in the mean-field
rate equation (\ref{reclv}).
As one should indeed expect on general grounds \cite{hkj01}, this continuous
active-to-absorbing phase transition is governed by the directed-percolation
universality class.
This fact can be established formally by starting from the Doi--Peliti action, 
and reducing it to Reggeon field theory near the predator extinction threshold
\cite{mgt07, uct12}.
Heuristically, the prey population almost fills the entire lattice in this situation;
consequently, the presence of $B$ particles sets no constraint on the 
predation process $A + B \to A + A$, which reduces to the branching
reaction $A + A \to A$.
Along with predator death $A \to \emptyset$ and the population-limiting
pair fusion reaction $A + A \to A$ that is in fact equivalent to local site
occupation restrictions, one recovers the fundamental processes of directed
percolation.
Extensive Monte Carlo simulations confirm directed-percolation critical
exponents at the predator extinction transition \cite{mgt07, st16, dmpt18}, 
including the associated critical aging scaling behavior for the two-time 
density auto-correlations \cite{hp10, uct14}.

\subsection{Clusters and spiral patterns in cyclic competition games}

Extension of two-species models to systems with multiple reactants in general 
increases the complexity of the problem tremendously \cite{dmpt18}. 
Nevertheless, simpler sub-structures can be amenable to full theoretical 
analysis.
For example, on occasion elementary competition cycles are present in food
networks.
As a final illustration of the often decisive role of spatial correlations in
stochastic chemical reactions or population dynamics, we therefore consider
the following cyclic interaction scheme involving three species subject to
Lotka--Volterra predation: $A + B \to A + A$ with rate $\lambda_A$;
$B + C \to B + B$ with rate $\lambda_B$; and 
$A + C \to C + C$ with rate $\lambda_C$, akin to the rock-paper-scissors
game \cite{hs98}.
These elementary replacement processes conserve the total particle number 
$N = N_A + N_B + N_C$ and hence density $\rho = a(t) + b(t) + c(t)$; of
course this is also true on the level of the coupled rate equations
\begin{eqnarray}
  &&\frac{\partial a(t)}{\partial t} = a(t) 
  \left[ \lambda_A \, b(t) - \lambda_C \, c(t) \right] \, , \quad
  \frac{\partial b(t)}{\partial t} = b(t) 
  \left[ \lambda_B \, c(t) - \lambda_A \, a(t) \right] \, , \nonumber \\
  &&\frac{\partial c(t)}{\partial t} = c(t) 
  \left[ \lambda_B \, a(t) - \lambda_B \, b(t) \right] \, ;
\label{rerps}
\end{eqnarray}
aside from the three absorbing states $(a,b,c) = (\rho,0,0)$, $(0,\rho,0)$,
and $(0,0,\rho)$, which are linearly unstable under the mean-field dynamics
(\ref{rerps}), yet represent the sole possible final configurations in any finite 
stochastic realization, there exists one neutrally stable reactive coexistence 
state:
$a(\infty) = \rho \, \lambda_B / (\lambda_A + \lambda_B + \lambda_C)$,
$b(\infty) = \rho \, \lambda_C / (\lambda_A + \lambda_B + \lambda_C)$,
$c(\infty) = \rho \, \lambda_A / (\lambda_A + \lambda_B + \lambda_C)$.
The linear stability matrix at this fixed point reads
\begin{equation*}
  {\bf L} = \frac{\rho}{\lambda_A + \lambda_B + \lambda_C}
  \begin{pmatrix} 
  \ 0 & \ \lambda_A \, \lambda_B & - \lambda_B \, \lambda_C \ \\ 
  - \lambda_A \, \lambda_C \ & \ 0 & \ \lambda_B \, \lambda_C \\
  \ \lambda_A \, \lambda_C & - \lambda_A \, \lambda_B \ & \ 0 
\end{pmatrix} ;
\end{equation*}
one of its eigenvalues is zero reflecting the conservation law $\rho =$const.,
the other two are purely imaginary, 
$\epsilon_\pm = \pm i \rho \, \sqrt{\lambda_A \, \lambda_B \, \lambda_C /
  (\lambda_A + \lambda_B + \lambda_C)}$, indicating undamped population
oscillations.

\begin{figure}[ht]
\centerline{
\subfigure[]{\includegraphics[width = 4cm, height = 4cm]{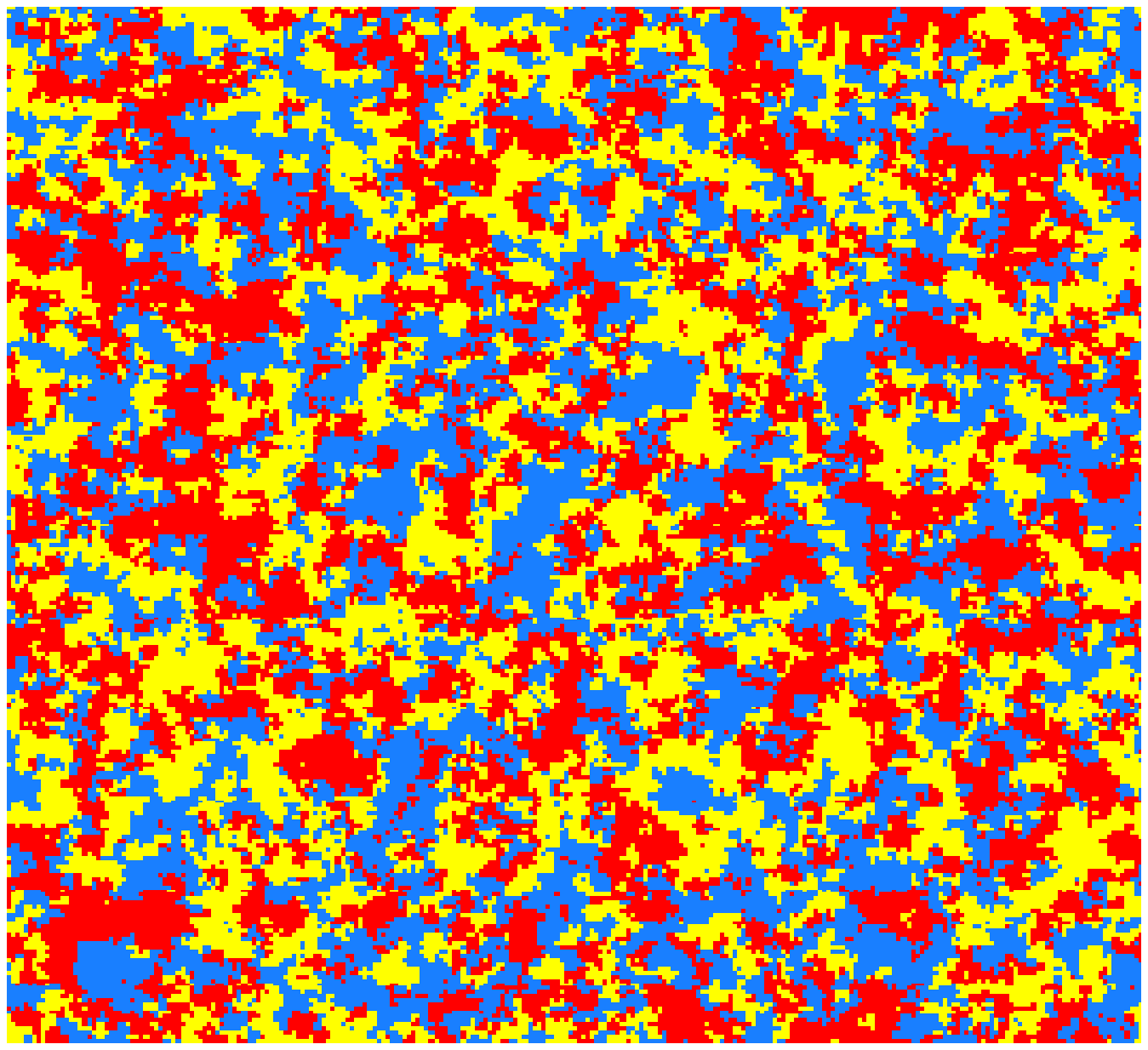}} 
\qquad 
\subfigure[]{\includegraphics[width = 4cm, height = 4cm]{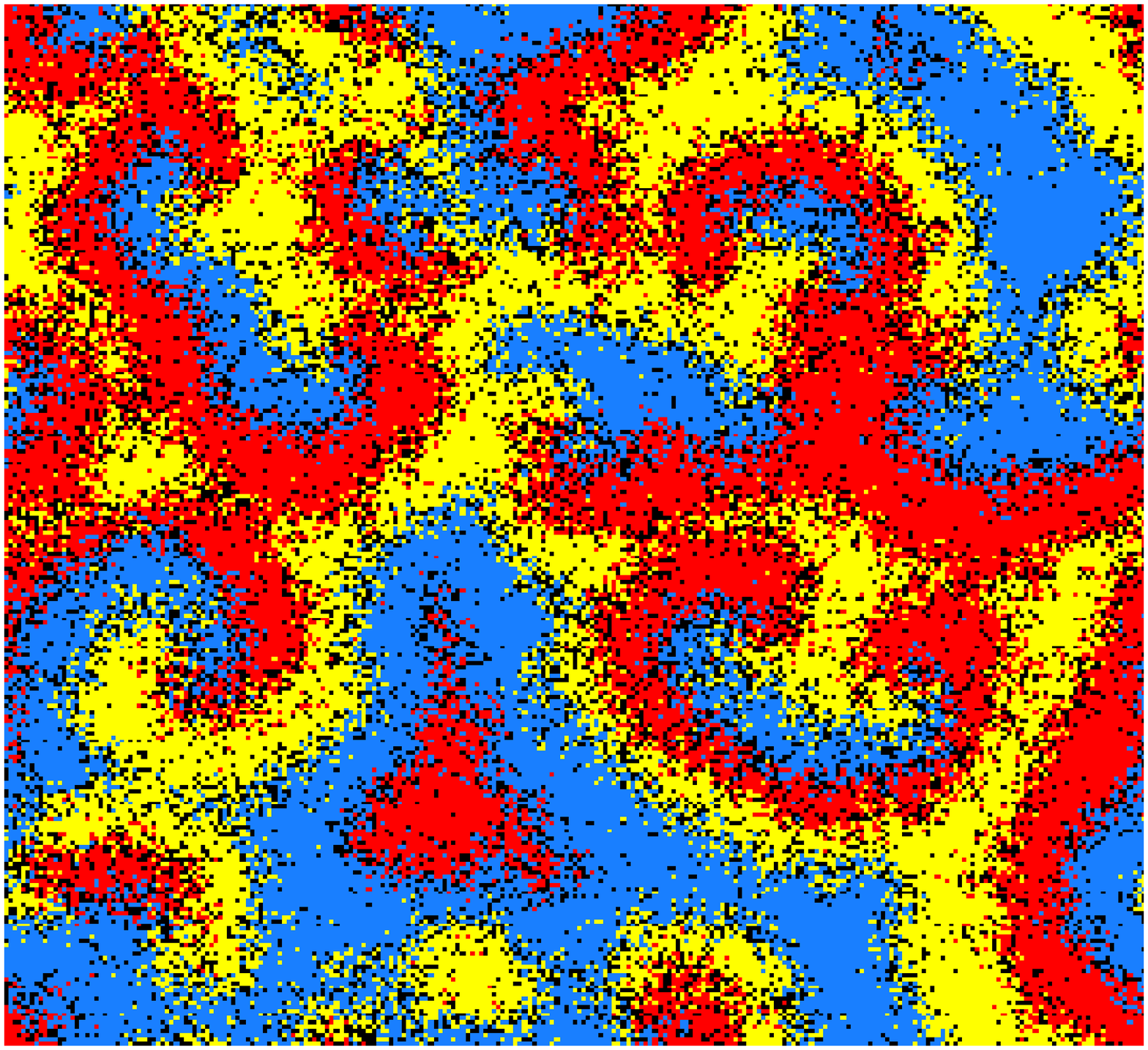}}}
\caption{Snapshots of a stochastic cyclic competition models on square
  lattices with $256 \times 256$ sites, periodic boundary conditions, site
  occupation number exclusion, and random initial particle placement:
  (a) cyclic Lotka--Volterra or rock-paper-scissors model with conserved total
  particle number;
  (b) May--Leonard model with separate predation and reproduction 
  processes.
  Sites occupied by $A$, $B$, and $C$ particles are respectively color-coded
  in red, yellow, and blue; empty spaces in black.
  [Figures reproduced with permission from (a) Q. He, M. Mobilia, and 
   U.~C. T\"auber, \emph{Phys. Rev. E} {\bf 82}, 051909-1--11 (2010);
   DOI: 10.1103/PhysRevE.82.051909; copyright (2010) by The American
   Physical Society; (b) Q. He, M. Mobilia, and U.~C. T\"auber,
   \emph{Eur. Phys. J. B} {\bf 82}, 97--105 (2011); 
   DOI: 10.1140/epjb/e2011-20259-x; copyright (2011) by EDP Sciences.]}  
\label{ccfig}
\end{figure}

Akin to the Lotka--Volterra two-species competition model discussed 
previously, one might thus anticipate traveling wave structures.
Yet Monte Carlo simulations of this rock-papers-scissors model on 
two-dimensional lattices that are sufficiently large to prevent extinction events
and stabilize the three-species coexistence state, one merely observes weakly
fluctuating clusters containing particles of the same type \cite{pa08, hmt10},
as shown in Fig.~\ref{ccfig}(a).
Upon initializing the system with a random spatial distribution, transient
population oscillations appear, with a characteristic frequency that differs
considerably from the mean-field prediction, but they are strongly damped,
at variance with the results from the rate equation analysis.
These findings are perhaps even more curious given in light of the fact that in
the extreme asymmetric limit $\lambda_A \gg \lambda_B, \lambda_C$, the
rock-paper-scissors system effectively reduces to the Lotka--Volterra model
with predators $A$, prey $B$, and abundant, essentially saturated $C$ 
population with $c(\infty) \approx \rho$. 
Indeed, on the mean-field level, one finds to leading order 
$a(\infty) \approx \rho \, \lambda_B / \lambda_A$, 
$b(\infty) \approx \rho \, \lambda_C / \lambda_A$, and $c(\infty) \approx 
 \rho \left[ 1 - (\lambda_B + \lambda_C) / \lambda_A\right]$.
We may thus replace $c(t) \approx \rho$ in the rate equations (\ref{rerps}),
which takes us to \eref{reqlv} with effective predator death rate 
$\mu = c(\infty) \, \lambda_C$ and prey reproduction rate 
$\sigma = c(\infty) \, \lambda_B$.
Lattice simulations of strongly asymmetric rock-paper-scissors model display
the characteristic Lotka--Volterra spreading activitiy fronts and persistent
oscillatory kinetics, and hence confirm this picture \cite{htz12}.

In the May--Leonard model variant of cyclic competition, predation and birth
reactions are explicitly separated: $A + B \to A$, $A \to A + A$ (and 
cyclically permuted) with independent rates \cite{hs98}.
In this coupled reaction scheme, the total particle number is not conserved
anymore, but still permits an active three-species coexistence state.
In the absence of intra-species competition, i.e., for infinite carrying
capacities, one eigenvalue of the associated linear stability matrix ${\bf L}$
is real and negative, corresponding to an exponentially relaxing mode.
The other two eigenvalues are purely imaginary, signifying the existence of
undamped oscillations.
Monte Carlo simulations of the May--Leonard model on two-dimensional 
lattices show the spontaneous formation of spiral patterns with alternating
$A$, $B$, $C$ species in sequence \cite{rmf07, pa08, dmpt18}.
These spiral structures become especially clear and pronounced if in addition
to nearest-neighbor hopping, particle exchange $A \leftrightarrow B$ etc. is
implemented, as illustrated in Fig.~\ref{ccfig}(b) \cite{hmt11}.

On the mean-field rate equation level, one may argue the relaxing eigenmode
to be irrelevant for the long-time dynamics of this system; confining the
subsequent analysis to the reactive plane spanned by the remaining 
oscillating modes, a mapping to the complex Ginzburg--Landau equation has
been achieved \cite{rmf08}, which is known to permit spiral spatio-temporal
structures in certain parameter regimes \cite{ak02}.
A consistent treatment of intrinsic reaction noise for this model based again
on the Doi--Peliti formalism however results in a more complex picture.
Since the relaxing mode is randomly driven by non-linear fluctuations that
involve the persistent oscillatory eigenmodes, it can only be integrated out in
a rather restricted parameter range that allows for distinct time scale
separation. In general, therefore the May--Leonard model is aptly described 
by a coupled set of three Langevin equations with properly constructed 
multiplicative noise \cite{st17}.

\section{Summary and Concluding Remarks}

In this chapter, we have given an overview over crucial fluctuation and 
correlation effects in simple spatially extended particle reaction systems that 
originate from the underlying stochastic kinetics; these very same models also
pertain to the basic phenomenology for infectious disease spreading in 
epidemiology and the dynamics of competing populations in ecology.
As befits the theme of this volume, the essential dynamical features of the 
paradigmatic models discussed here require analysis beyond the standard
textbook fare that mostly utilizes coupled deterministic rate equations, which
entail a factorization of correlation functions into powers of the reactants' 
densities, thereby neglecting both temporal and spatial fluctuations.
 
Spontaneous death-birth reactions augmented with population-limiting fusion 
constitute the fundamental simple epidemic process.
The ensuing population extinction threshold exemplifies a continuous 
non-equilibrium phase transition between active and absorbing states which
is naturally governed by large and long-range fluctuations.
It is generically characterized by the critical scaling exponents of directed
percolation, which assume non-trivial values in dimensions below the upper
critical dimension $d_c = 4$.

Binary (or triplet) annihilation reactions generate depletion zones in 
dimensions $d \leq d_c = 2$ $(1)$; the particle anti-correlations cause a
drastic slowing-down of the reaction kinetics in the diffusion-limited regime.
For pair annihilation involving two distinct species, spatial segregation in
dimensions $d \leq d_s = 4$ confines the reactions to the interfaces 
separating the diffusively coarsening domains.
In the special case of precisely equal species densities, this further diminishes
the overall reaction activity, and considerably decelerates the asymptotic
algebraic decay.

In stochastic, spatially extended variants of the classical Lotka--Volterra model
for predator-prey competition, the intrinsic demographic noise causes and
stabilizes spreading activity fronts in the species coexistence phase; their 
quantitative properties are strongly affected by the stochastic fluctuations and
spatial correlations self-generated by the fundamental reaction kinetics in the
system.
Furthermore, the implementation of finite local carrying capacities for the
prey population, mimicking limitations in their food resources, generates an
extinction threshold for the predator species.
Finally, we have elucidated some basic features in cyclic competition models
that involve three particle species, and related the resulting spatial structures, i.e., population clusters for direct rock-paper-scissors competition and 
dynamical spirals in the May--Leonard model, to the presence and absence,
respectively, of a conservation law for the total particle number. 

In more complex systems that involve a larger number of reactants, many of
these fluctuation-dominated features may of course be effectively averaged
out or become inconspicuous on relevant length and/or time scales.
However, even when important qualitative properties such as the topology of
the phase diagram or the essential functional form of the time evolution remain adequately captured by mean-field rate equations, internal fluctuations
as well as emerging spatial and temporal patterns or even short-lived
correlations may well drastically modify the numerical values of effective 
reaction rates and transport coefficients. 

Two case studies from the author's research group may illustrate this point:
(I) In ligand-receptor binding kinetics on cell membranes or surface plasmon
resonance devices that are commonly utilized to measure reaction affinities,
repeated ligand rebinding to nearby receptor locations entails temporal 
correlations that persist under diffusive dynamics and even advective flow; the
correct reaction rate values may consequently deviate drastically from
numbers extracted by inadequate fits to straightforward rate equation kinetics
\cite{crft16}.
(II) Circularly spreading wave fronts of `killer' as well as resistant bacteria
strains induce the formation of intriguing patterns in an effectively 
two-dimensional synthetic \emph{E. coli} micro-ecological system,
emphasizing the important role of spatial inhomogeneities and local
correlations \cite{dmcstjb17}.
A sound understanding of the potential effects of fluctuations and correlations
is thus indispensable for proper theoretical modeling and analysis as well as
quantitative interpretation of experimental data.

There are of course other crucial fluctuation and correlation effects on
chemical reactions and population dynamics that could not be covered in this
chapter \cite{dmpt18}.
Prominent examples are the strong impact of intrinsic noise on extinction  probabilities and pathways in finite systems \cite{am17}; and novel 
phenomena related to extrinsic random influences or quenched internal 
disorder \cite{tv06}.
These topics and many more are addressed elsewhere in this book.

\section*{Acknowledgments}
I would like to sincerely thank my collaborators over the past 25 years 
for their invaluable insights and crucial contributions to our joint research on 
stochastic reaction-diffusion and population dynamics, especially: 
Timo Aspelmeier, Nicholas Butzin, John Cardy, Jacob Carroll, Sheng Chen,
Udaya Sree Datla, Olivi\'er Deloubri\`ere, Ulrich Dobramysl, 
Kim Forsten-Williams, Erwin Frey, Ivan Georgiev, Yadin Goldschmidt, 
Manoj Gopalakrishnan, Qian He, Bassel Heiba, Henk Hilhorst, 
Haye Hinrichsen, Martin Howard, Hannes Janssen, Weigang Liu, 
Jer\^ome Magnin, William Mather, Mauro Mobilia, Michel Pleimling, 
Matthew Raum, Beth Reid, Gunter Sch\"utz, Franz Schwabl (deceased),
Shannon Serrao, Steffen Trimper, Ben Vollmayr-Lee, Mark Washenberger,
Fr\'ed\'eric van Wijland, and Royce Zia.

This research was in part sponsored by the US Army Research Office and was 
accomplished under Grant Number W911NF-17-1-0156. 
The views and conclusions contained in this document are those of the author
and should not be interpreted as representing the official policies, either 
expressed or implied, of the Army Research Office or the US Government.
The US Government is authorized to reproduce and distribute reprints for 
Government purposes notwithstanding any copyright notation herein.

\end{document}